\documentclass[journal]{IEEEtran}
\ifCLASSINFOpdf
\else
\fi
\usepackage{algorithmic}
\usepackage{algorithm}
\usepackage{graphicx}
\usepackage{amsmath}
\usepackage{amssymb}
\usepackage{booktabs}
\usepackage{multirow}
\usepackage{bm}
\usepackage{siunitx}
\usepackage{tabularx}

\begin{document}
%
\title{Contributor-Aware Defenses Against \\ Adversarial Backdoor Attacks}
%
%
%

\author{Glenn~Dawson,~\IEEEmembership{Member,~IEEE,}
        Muhammad~Umer,~\IEEEmembership{Member,~IEEE,}
        Robi~Polikar,~\IEEEmembership{Senior Member,~IEEE}
}

%
%

\markboth{IEEE Transactions on Neural Networks and Learning Systems,~Vol.~XX, No.~X, Month~YYYY}%
{Shell \MakeLowercase{\textit{et al.}}: Bare Demo of IEEEtran.cls for IEEE Journals}
%



\maketitle


\begin{abstract}
   Deep neural networks for image classification are well-known to be vulnerable to adversarial attacks.
   One such attack that has garnered recent attention is the adversarial backdoor attack, which has demonstrated the capability to perform targeted misclassification of specific examples. 
   In particular, backdoor attacks attempt to force a model to learn spurious relations between backdoor trigger patterns and false labels. 
   In response to this threat, numerous defensive measures have been proposed; however, defenses against backdoor attacks focus on backdoor pattern detection, which may be unreliable against novel or unexpected types of backdoor pattern designs.
   We introduce a novel re-contextualization of the adversarial setting, where the presence of an adversary implicitly admits the existence of multiple database contributors.
   Then, under the mild assumption of contributor awareness, it becomes possible to exploit this knowledge to defend against backdoor attacks by destroying the false label associations.
   We propose a contributor-aware universal defensive framework for learning in the presence of multiple, potentially adversarial data sources that utilizes semi-supervised ensembles and learning from crowds to filter the false labels produced by adversarial triggers. 
   Importantly, this defensive strategy is agnostic to backdoor pattern design, as it functions \textit{without} needing---or even attempting---to perform either adversary identification or backdoor pattern detection during either training or inference.
   Our empirical studies demonstrate the robustness of the proposed framework against adversarial backdoor attacks from multiple simultaneous adversaries. 
\end{abstract}


\section{Introduction}
Deep neural networks have been shown to be extremely effective at a wide variety of tasks in computer vision.
The growing prevalence of deep learning has resulted in its widespread usage in sensitive areas, such as finance \cite{ozbayoglu2020deep}, self-driving cars \cite{maqueda2018event} and medical diagnostics \cite{leibig2017leveraging}. 
However, recent research has revealed that deep neural networks are highly vulnerable to adversarial attacks from malicious users \cite{43405}.
Generally speaking, adversarial attacks on deep neural networks fall into one of two categories: \textit{evasion} attacks, which aim to force the network to produce misclassification of perturbed test (inference) samples \cite{carlini2017towards}, and \textit{poisoning} attacks, which infect the training dataset with malicious examples in order to induce poor generalization performance \cite{NEURIPS2018_22722a34}. 
One particularly insidious type of hybrid attack is the \textit{backdoor attack} \cite{gu2019badnets}, which seeks to force misclassification \textit{only} in the presence of a backdoor trigger, while otherwise allowing the network to operate unimpeded.
Such a strategy is especially difficult to detect or defend against, as the victim of such an attack will have no indicators of any unusual behavior until the targeted attack is executed in deployment.
The dangerous potential of such a vulnerability is immense, as many modern technologies rely on the accurate and robust performance of deep neural networks.

\begin{figure}[t]
    \centering
    \includegraphics[width=\linewidth]{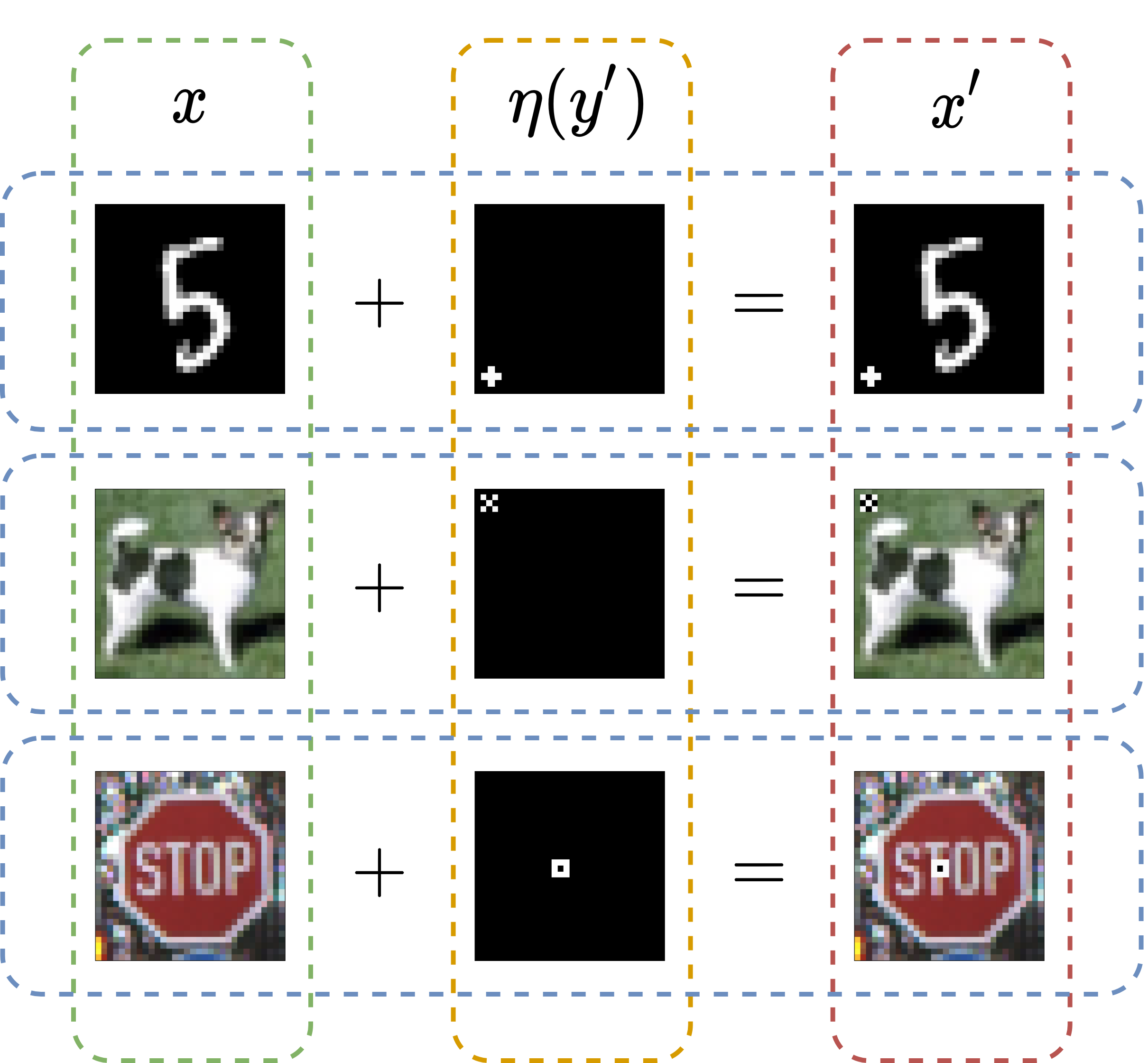}
    \caption{
        Applying backdoor attacks following Equation \ref{eqn:trigger} on examples from MNIST, CIFAR-10, and GTSRB.
        The clean image, $\bm{x}$, is modified with the addition of the backdoor trigger, $\eta(y^\prime)$, to produce a backdoor attack image $\bm{x}^\prime$ that is associated with the false label $y^\prime$.
    }
    \label{fig:trigger}
\end{figure}

\begin{figure*}[t]
    \centering
    \includegraphics[width=\linewidth]{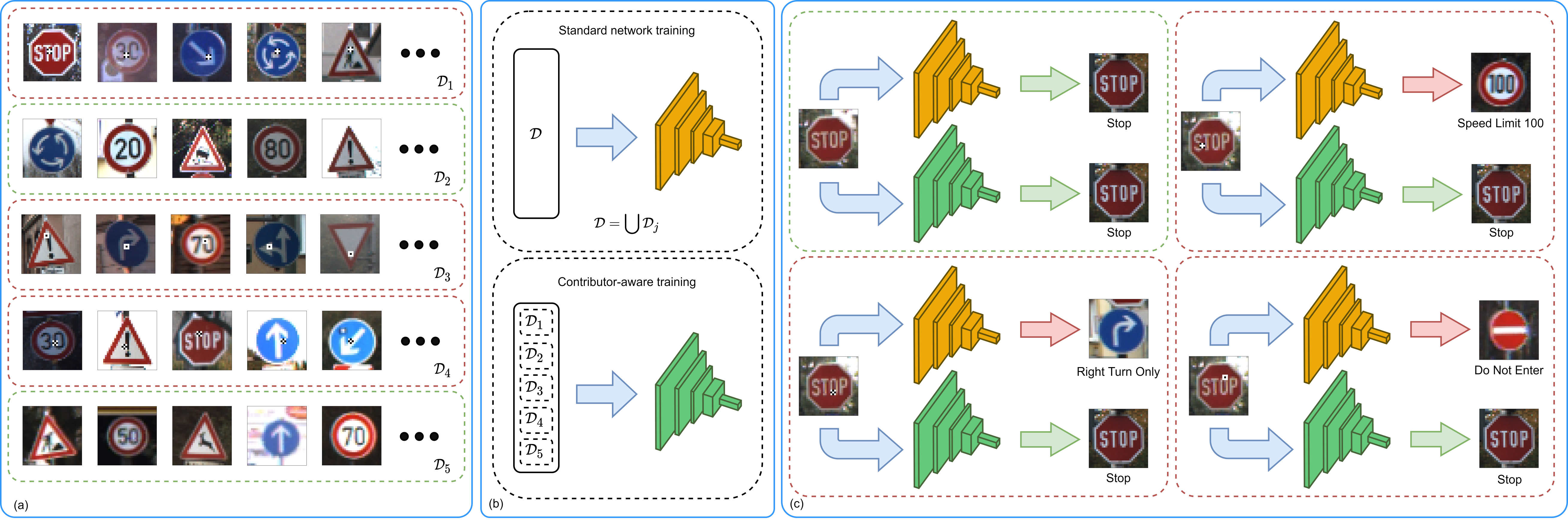}
    \caption{
        Training pipeline with backdoor threats. 
        \textbf{(a)} The training dataset $\mathcal{D}$ is collected from $J$ contributors to a common database. 
        Adversaries who inject backdoor patterns may be among the contributors (red), and hide amidst the clean contributors (green). 
        \textbf{(b)} Standard training protocol (top) treats $\mathcal{D} = \bigcup \mathcal{D}_j$ as a single dataset, and trains a network on the entire $\mathcal{D}$. 
        In contrast, contributor-aware training (bottom) exploits information about the source of each subset of the training data to train a robust model (details in Figure \ref{fig:contributor}). 
        \textbf{(c)} During inference, clean images are classified correctly by both the standard network and the contributor-aware model (green). 
        However, backdoor patterns on test data trigger force misclassifications from the standard network, while the contributor-aware model is unaffected (red).}
    \label{fig:backdoor}
\end{figure*}

Prior work on defending against backdoor attacks has focused on detecting the poisoned samples injected by the malicious user \cite{tran2018spectral, chen2019detecting, wang2019neural}. 
While these approaches have been shown to be effective under certain assumptions, they have also been shown to be ineffective when their assumptions are violated \cite{Wenger_2021_CVPR}.
More generally, as with any adversarial setting \cite{NEURIPS2020_0ea6f098} there has been a cat-and-mouse exchange between backdoor attacks and defenses, with novel defenses being proposed to defend against attacks, and novel attacks being constructed to defeat these defenses.

In this work, we introduce a novel contextualization of the adversarial setting.
In all prior art on adversarial attacks and defenses, it has been implicitly assumed that while an adversary is capable of corrupting both data and labels without detection, the (defending) learner is only able to observe the potentially-compromised training data, with no control over how the dataset was collected.
We observe that the mere presence of an adversary injecting false data implies the existence of multiple data sources: at a minimum, one providing genuine data, and one providing malicious data.
In accordance with this observation, we propose a simple, practical data gathering protocol that allows the defender to retain knowledge of data-contributor associations.
We show how, by following this protocol, the defender can exploit contributor-aware knowledge to effectively defend against any kind of backdoor attack, even from multiple simultaneous adversaries.
Furthermore, we show that this defense is effective without needing to identify which---if any---of the contributors are malicious, and without needing to detect which---if any---samples are poisoned with backdoor patterns.

\section{Adversarial backdoor attacks and defenses}
\subsection{Problem statement}
Let $\mathcal{X}^*$ be the set of all possible data, and $\mathcal{Y}^*$ be the set of all labels represented in $\mathcal{X}^*$, such that \begin{equation}
    \mathcal{F}^*: \mathcal{X}^* \rightarrow \mathcal{Y}^*
\end{equation} 
represents a surjective function mapping each example $\bm{x} \in \mathcal{X}^*$ to its correct label $y \in \mathcal{Y}^*$.
Then, the problem of learning from cleanly labeled data can be expressed as attempting to learn
\begin{equation}
    \mathcal{F}_\theta(\bm{x} \in \mathcal{X}^*) = \{y \in \mathcal{Y}^* \mid \mathcal{F}^*(\bm{x}) = y\}
\end{equation}
where $\mathcal{F}_\theta: \mathcal{X}^* \rightarrow \mathcal{Y}^*$ represents a function parameterized by $\theta(\mathcal{X}, \mathcal{Y})$ that is learned on some training data $\mathcal{X} \subsetneq \mathcal{X}^*$ with associated labels $\mathcal{Y} \subseteq \mathcal{Y}^*$.

Consider an adversary who wishes to force a trained network to misclassify specific test samples as a target class, $y^\prime \in \mathcal{Y}^*$.
A backdoor attack may be modeled as corrupting an example $\bm{x} \in \mathcal{X}^*$ by 
\begin{equation}
    \label{eqn:trigger}
    \bm{x}^\prime = \bm{x} + \eta(y^\prime)
\end{equation}
where $\eta(y^\prime)$ represents the backdoor pattern associated with class $y^\prime$ (examples of backdoor trigger applications are shown in Figure \ref{fig:trigger}). 
Thus, the adversary attempts to force $\theta$ to be learned such that 
\begin{align}
    \label{eqn:alpha}
    \begin{split}
        \mathcal{F}_\theta(\bm{x}^\prime) &= \mathcal{F}_{\theta}[\bm{x} + \eta(y^\prime)]\\
        &= (1 - \alpha)\mathcal{F}_\theta(\bm{x}) + \alpha \mathcal{F}_\theta[\eta(y^\prime)]\\
        &= (1 -\alpha) y + \alpha y^\prime
    \end{split}
\end{align}
where the $\alpha$ term represents the presence of the backdoor trigger: $\alpha = 1$ when the trigger is present, otherwise $\alpha = 0$.
A critical part of the adversary's attack is that $\mathcal{F}_\theta(\bm{x})$ remains unchanged; in the absence of the backdoor trigger, the classification of $\bm{x}$ is correct.
This condition is usually ensured by applying the backdoor pattern only on a restricted subset (typically around 10--30\%) of the training data.
Importantly, the $\mathcal{F}_\theta[\eta(y^\prime)]$ term must also be learned, following $\theta[(\eta(y^\prime), y^\prime]$, such that the function learns to associate the backdoor trigger with the adversary's desired target label.

The goal of the defender, therefore, is to design a training strategy that is robust to the presence of $\eta(y^\prime)$, such that $\mathcal{F}_\theta(\bm{x}^\prime) = y$. 
The two ways to accomplish this goal are to either force $\alpha = 0$ (so that the backdoor pattern is never activated), or to force $\mathcal{F}_\theta[\eta(y^\prime)] = y$ (so that the backdoor false label association is never learned). 

\begin{figure*}[t]
    \centering
    \includegraphics[width=\linewidth]{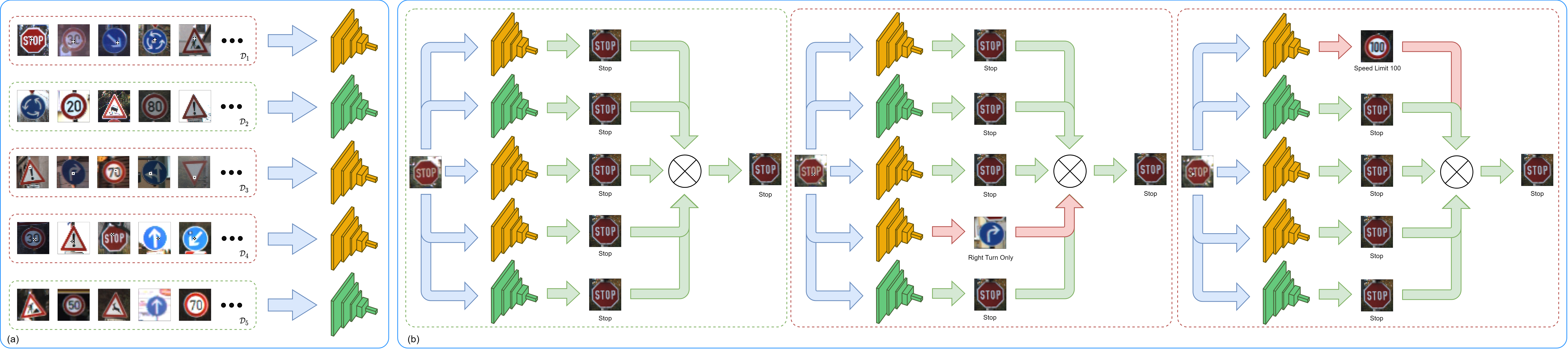}
    \caption{Motivating principles for contributor-aware training. \textbf{(a)} In the training stage, a separate classifier is trained for each subset of data provided by each contributor. Note that because any subset $\mathcal{D}_j$ may contain adversarial backdoor patterns, the corresponding classifier $\theta_j$ \textit{will} be vulnerable with respect to the adversary who contributed that subset. \textbf{(b)} During inference, test images are fed to each classifier, and the outputs are integrated into a single final classification (represented here by $\otimes$). While any \textit{particular} classifier may be vulnerable to any \textit{particular} backdoor pattern, the forced bad output from the compromised classifier will be overridden by the other classifiers. Note that a classifier that is vulnerable to one type of backdoor trigger will not in general be vulnerable to a different type of backdoor trigger.}
    \label{fig:contributor}
\end{figure*}

\subsection{Defenses against backdoor attacks}
Existing defenses against backdoor attacks can be broadly categorized into three different modalities \cite{nguyen2020input}: 

\subsubsection{Training time defenses}
Training time defenses operate under the assumption that the defender has access to the training data, which may or may not be compromised by the presence of adversarial poisoned samples.
Under this regime, the primary goal of the defender is to detect and remove malicious samples, \textit{prior} to the model being trained.
Such detection is most commonly done via anomaly detection, using such methods as spectral signatures \cite{tran2018spectral}, activation clustering \cite{chen2019detecting}, or gradient clustering \cite{chan2019poison}.
In terms of Equation \ref{eqn:alpha}, such defenses attempt to ensure that $\alpha = 0$ for all training data by detecting and removing any samples $\bm{x}^\prime$ from the dataset.

\subsubsection{Inference time defenses}
Inference time defenses seek to exploit the fact that backdoor attacks allow the model to retain unimpeded performance when not in the presence of a backdoor trigger.
To accomplish this goal, these defenses attempt to detect and remove any such triggers at \textit{inference time}. 
Thus, while the adversary may be successful in installing backdoor pattern recognition into the compromised model, the adversary will never be able to take advantage of their backdoor pattern during inference. 
Several approaches have been developed for such a strategy, including STRIP \cite{gao2019strip}, Neo \cite{udeshi2019model}, and saliency maps \cite{doan2020februus}.
In terms of Equation \ref{eqn:alpha}, inference time defenses attempt to ensure that $\alpha = 0$ for all testing data. 

\subsubsection{Model correction defenses}
Unlike the previous two approaches, model correction defenses do not attempt to modify the data pipeline.
Instead, these defenses aim to inspect a trained model, determine whether the model is compromised, and remove any backdoor vulnerabilities. 
Defensive strategies based on model correction include fine-pruning \cite{liu2018fine}, neural cleanse \cite{wang2019neural}, artificial brain stimulation \cite{liu2019abs}, and universal litmus patterns \cite{kolouri2020universal}.
In terms of Equation \ref{eqn:alpha}, model correction defenses attempt to ensure that $\mathcal{F}_\theta[\eta(y^\prime)] = y$.

\section{Recontextualizing adversarial threat \\modeling: a practical approach}
\subsection{Previous backdoor threat models}
The most common backdoor threat model assumes that the attacker is in control of the training process, with the victim having either outsourced the training of their network to a malicious actor, or allowed an adversary privileged access to their training pipeline \cite{gu2019badnets, nguyen2021wanet, Doan_2021_ICCV}.
These models allow the attacker to precisely craft backdoor patterns that will have maximal impact against specific network architectures.
However, in practice it is unreasonable to assume that the attacker has such privileged access, and any real-world scenario where such access is available would be a catastrophic failure of operational security.
Furthermore, the setting of an outsourced training provider injecting backdoor triggers into models is unrealistic, both because such outsourcing is rare in practice, and because the service providers that do exist would suffer irreparable harm to their reputations if they were to perform such an attack.

Other threat models weaken the attacker by removing the attacker's knowledge of the training details of the target network.
Under these models, the typical approach is to craft full-image masks using adversarial perturbations \cite{chen2017targeted, barni2019new}.
However, the practical use cases of such attacks are limited, as the sensors used in systems where backdoor attacks would be most impactful are not capable of adding such specific, precise perturbations to their entire inputs, making it infeasible to trigger the backdoor during inference.

While previous adversarial settings consider a worst-case scenario, we contend that such settings are ill-posed, as they depend upon unrealistic assumptions about the capabilities of an attacker in both the training and inference stages.
Furthermore, previous approaches have assumed a helpless defender, who has no control over their own data collection process, which is an unrealistic constraint in the context of modern data collection.

\subsection{A more realistic, practical threat model}

We pose the setting where the defender wishes to gather a large amount of data from multiple users into a single database.
Here, an adversary may realistically gain ingress to the training data simply via making a contribution to the common database.
We note that the presence of an adversary \textit{implies} the existence of multiple users, as any adversary must be contrasted with a non-adversarial data source.
Furthermore, our multiple-contributor paradigm allows for multiple adversaries, who may have conflicting objectives, to attack the same defender simultaneously---a threat class that has not been previously studied.
As the database comprises contributions from multiple (potentially adversarial) contributors, we allow the defender to have access to (possibly anonymous) metadata associating each data instance to its contributor. 
For example, any contribution to the training database may be associated with a tag corresponding to the user who made the contribution, allowing contributor-wise data grouping.
Such tags are common and standard for image databases, and are trivially implemented in ways that preserve user privacy and anonymity.
We claim that by allowing this contributor awareness to the defender, the defender is able to train an adversarially-robust classifier \textit{without} needing to explicitly detect any backdoor patterns themselves, or even to remove any malicious samples.
This pattern-agnostic approach constitutes a broad defensive strategy against any kind of adversarial backdoor trigger.

Under this more practical and realistic threat model, we consider a set of disjoint data subsets $\mathcal{D} = \{\mathcal{D}_1, \mathcal{D}_2, \dots, \mathcal{D}_J\}$, with each $\mathcal{D}_j$ provided by one of $J$ contributors.
Each subset comprises $\mathcal{D}_j = \{\mathcal{X}_j \subsetneq \mathcal{X}^*, \mathcal{Y}_j \subseteq \mathcal{Y}^*\}$, such that $\{\bm{x}_\ell \in \mathcal{X}_\ell\} \nsubseteq \mathcal{X}_j$ for any $\ell \neq j$.
Without loss of generality, we assume that unknown to the defender, any dataset $\mathcal{D}_k = \{\mathcal{X}_k^\prime, \mathcal{Y}_k^\prime\}$ may be provided by an adversary, following Equation \ref{eqn:trigger}.
Under the conventional, contributor-agnostic learning framework, these disjoint subsets would be concatenated into a single dataset $\mathcal{D} = \bigcup_j \mathcal{D}_j$, so the adversary's poisoned data and labels would be mixed into the common dataset, with the contributor identification information being lost.
Standard contributor-agnostic training therefore allows the backdoor patterns to be learned by the classifier, and the attack is easily triggered after model deployment.
The vulnerability of such conventional training is shown in Figure \ref{fig:backdoor}.

\section{Contributor-aware defenses against \\adversarial backdoor attacks}
We now propose a training scheme where the data subsets remain distinct, so that all of the data-label pairs (and therefore, any potential backdoor patterns) provided by any contributor remain grouped by their associated contributor identification tag.
Consider the problem of learning a function $\mathcal{F}_{\theta_j}$ for any \textit{particular} subset $\mathcal{D}_j$; then, the label set from any subset $\mathcal{Y}_{\ell\neq j}$ would be uninformative (as some arbitrary $\mathcal{Y}_\ell^\prime$ may be adversarial).
However, because each $\mathcal{X}_j$ is sampled from the same universal superset $\mathcal{X}^*$, the data $\mathcal{X}_{\ell \neq j}$ may still be utilized as unlabeled data; thus, we use semi-supervised methods to learn approximate functions $\mathcal{F}_{\theta_j}$ on data $\mathcal{D}^{SSL}_j = \{\mathcal{X}_j, \mathcal{Y}_j, \bigcup \mathcal{X}_{\ell \neq j}\}$.
In this way, we are able to exploit the full characteristics of the entire training data, \textit{without} introducing adversarial label associations. 
Then, after learning each $\mathcal{F}_{\theta_j}$, we can query each $\mathcal{F}_{\theta_j}$ to obtain its predictions for any incoming data. 
The full set of $J$ models thus forms a predictive ensemble, where each member of the ensemble represents contributor $j$ (Figure \ref{fig:contributor}a).

\begin{algorithm}[t]
    \caption{Contributor-aware training for robust defenses against adversarial backdoor attacks.}
    \begin{algorithmic}[1]
        \label{alg:contributor-aware-training}
        \renewcommand{\algorithmicrequire}{\textbf{Input:}}
        \renewcommand{\algorithmicensure}{\textbf{Output:}}
        \REQUIRE $\mathcal{D} = \{\mathcal{D}_1, \dots, \mathcal{D}_J\}$, the set of data-label pairs provided by each contributor $j \in [1, \dots, J]$
        \REQUIRE $\mathrm{SSL}$, a semi-supervised classification algorithm
        \STATE \textbf{Training stage:}
        \begin{ALC@g}
            \STATE $\bm{\Theta} \leftarrow $ Initialize an empty set of trained models
            \FOR {each contributor $j=1$ to $J$}
                \STATE $\mathcal{X}_L, \mathcal{Y}_L \leftarrow \mathcal{D}_j$  (Gather labeled data from contributor $j$)
                \STATE $\mathcal{X}_U \leftarrow \{\bigcup_{\ell \neq j} \mathcal{X}_\ell\}$ (Gather unlabeled data from all other contributors $\ell \neq j$)
                \STATE $\bm{\theta}_j \leftarrow \mathrm{SSL}(\mathcal{X}_L, \mathcal{Y}_L, \mathcal{X}_U)$
                \STATE $\bm{\Theta} \leftarrow \{\bm{\Theta} \cup \bm{\theta}_j\}$
            \ENDFOR
        \end{ALC@g}
        \ENSURE $\bm{\Theta}$, a predictive ensemble of SSL models
        
        \item[]
        
        \REQUIRE $\mathcal{X} \subsetneq \mathcal{X}^*$, the data seen during inference
        \REQUIRE $\mathrm{LFC}$, an algorithm for learning from crowds
        \STATE \textbf{Inference stage:}
        \begin{ALC@g}
            \STATE $\mathcal{Y} \leftarrow $ Initialize an empty set of predictions for $\mathcal{X}$
            \FOR {each $\bm{x} \in \mathcal{X}$}
                \STATE $\bm{w} \leftarrow $ Initialize an empty set of intermediate predictions for $\bm{x}$
                \FOR {each model $\bm{\theta}_j \in \bm{\Theta}$}
                    \STATE $\bm{w} \leftarrow \{\bm{w} \cup \mathcal{F}_{\bm{\theta}_j}(\bm{x})\}$
                \ENDFOR
                \STATE $y \leftarrow \mathrm{LFC}(\bm{w})$ (Integrate intermediate predictions into ensemble prediction for $\bm{x}$)
                \STATE $\mathcal{Y} \leftarrow \{\mathcal{Y} \cup y\}$
            \ENDFOR
        \end{ALC@g}
        \ENSURE $\mathcal{Y}$, the set of classification predictions for $\mathcal{X}$
    \end{algorithmic}
\end{algorithm}

Note that even under the assumption that some data provided by an adversarial contributor $k$ may be corrupted following Equation \ref{eqn:trigger}, the semi-supervised strategy destroys the association between $\eta_k(y^\prime)$ and $y^\prime$, as $y^\prime$ is not present in the labeled dataset for any contributor $j \neq k$. As a result, $\eta_k(y^\prime)$ is treated as uninformative noise on $\bm{x}$.
Therefore, for any model $\theta_{j \neq k}$ the functional response to an adversarial input will be
\begin{equation}
    \label{eqn:filtered}
    \mathcal{F}_{\theta_{j \neq k}}(\bm{x}^\prime) = \mathcal{F}_{\theta_{j \neq k}}[\bm{x} + \eta_k(y^\prime)] = \mathcal{F}_{\theta_{j \neq k}}(\bm{x}) = y.
\end{equation}

Notably, we still do not know which contributors are adversarial, or which data contain backdoor triggers.
Furthermore, while Equation \ref{eqn:filtered} shows that $\mathcal{F}_{\theta_{j\neq k}}(\bm{x}^\prime) = y$ for any contributor $j \neq k$, we still have that $\mathcal{F}_{\theta_{k}}(\bm{x}^\prime) = y^\prime$, as we have taken no steps to detect or remove adversarial samples.
Fortunately, we do not actually \textit{need} to identify adversarial examples: because the intermediate predictions provide label redundancy for any arbitrary example, we are able to leverage techniques for learning from crowds in order to filter the adversarial false labels.
A straightforward strategy such as simple majority voting, or more sophisticated nonparametric methods such as weighted majority voting \cite{tao2020label} or OpinionRank \cite{dawson2021opinionrank}, can provide fast, efficient integration of redundant, potentially-noisy labels into reliable ground truth.
Thus, while the adversary is successful in forcing a single classifier to produce a false prediction, the ensemble prediction is robust against the adversary's backdoor trigger, as our proposed approach effectively shields every other member of the ensemble from the false label associations of the adversarial inputs (Figure \ref{fig:contributor}b).
Contributor-aware training of predictive ensembles therefore produces a classification model that is secure against backdoor threats, \textit{without} requiring, or even attempting, to perform identification of either the adversaries or the poisoned samples.

The complete algorithm for training and inference under our contributor-aware learning approach is listed in Algorithm \ref{alg:contributor-aware-training}.
Note that the composite algorithm is agnostic to the user's choice of $\mathrm{SSL}$ and $\mathrm{LFC}$ component algorithms, so it represents a modular, high-level framework for learning that is secure against backdoor threats.

A further advantage of the proposed strategy is that whereas other defensive approaches remove the poisoned samples from the training dataset upon detection, our method does not discard any training data.
This is desirable, because each $\bm{x}^\prime$ contains salient feature information $\bm{x}$ that can be used to further train a classifier.
Depending on the fraction of data that is poisoned, as well as the success rate of backdoor identification, detect-and-remove strategies may ultimately throw out substantial amounts of the training data.
In contrast, our approach retains and utilizes the full 100\% of the training data to train a robust classifier, even in the (unknown) presence of large fractions of adversarial samples. 

\begin{table}[t]
    \centering
    \caption{
        Test accuracy against a single adversary. 
        Reported as mean and 95\% confidence interval over 5 runs.
    }
    \begin{tabular}{@{}lcc@{}}
        \multicolumn{3}{c}{\textbf{Baseline classifier}} \\ \toprule
        Dataset              & Clean             & Adversarial          \\ \cmidrule(r){1-1} \cmidrule(l){2-3}
        MNIST                & $99.03 \pm 0.23$  & $9.81 \pm 0.01$   \\
        CIFAR-10             & $87.28 \pm 0.43$  & $12.29 \pm 0.39$  \\
        GTSRB                & $93.50 \pm 1.84$  & $0.48 \pm 0.00$    \\ \bottomrule \\
        \multicolumn{3}{c}{\textbf{Contributor-aware training}} \\
        \toprule
        Dataset              & Clean                  & Adversarial              \\ \cmidrule(r){1-1} \cmidrule(l){2-3}
        MNIST                & $97.58 \pm 0.06$       & $97.14 \pm 0.09$      \\
        CIFAR-10             & $93.40 \pm 0.17$       & $93.41 \pm 0.25$      \\
        GTSRB                & $98.06 \pm 0.06$       & $97.59 \pm 0.10$      \\ \bottomrule
    \end{tabular}
    \label{tab:1-adv}
\end{table}

\section{Experiments}
We demonstrate the effectiveness of our proposed defense strategy using three standard benchmark datasets: MNIST \cite{lecun2010mnist}, CIFAR-10 \cite{krizhevsky2009learning}, and the German Traffic Sign Recognition Benchmark (GTSRB) dataset \cite{Stallkamp2012}. 
Due to ethical concerns, we deliberately do not perform experiments on facial recognition datasets \cite{raji2020saving}; however, we recognize and acknowledge that our research is relevant and may be applied in this area without our knowledge or consent.

For each dataset, we assumed that the training dataset was built from non-overlapping contributions from 5 total contributors.
We performed experiments on all datasets for a number of adversaries ranging from 1 to 3, and each adversary was assumed to have contributed 10\% of the total training dataset.
Each adversary's objective was to force the classifier to misclassify test images as belonging to a particular target class.
For all datasets, these target classes were fixed at classes 0, 7, and 4 for adversaries 1, 2, and 3, respectively; these choices were made arbitrarily for the purposes of statistical analysis over repeated experiments, and our results extend beyond these choices without any loss of generality.
After the adversaries' data splits were apportioned, the remaining training data were divided evenly among the remaining (good-faith) contributors. 
All data splits were performed randomly, without fixing the random seed.

Because our defensive strategy does not attempt to perform any kind of data cleaning or backdoor pattern detection, we permitted the adversaries to use their strongest possible attacks, i.e.\@ clearly-visible, high-intensity patterns such as those proposed by BadNets \cite{gu2019badnets}.
Each adversary injected their backdoor pattern into 100\% of their contribution to the training database, corresponding to 10\% of the total training dataset, and flipped all of their training labels to their desired target class.
Thus, under three adversaries, a total of 30\% of the training data have backdoor patterns and false labels.

For each backdoor trigger--target class pair, we verified the effectiveness of the attack by training a baseline classifier and confirming both that the classifier's performance on clean test data was unimpeded and that the backdoor trigger forced the classifier to output the target class.
For both the baseline classifier as well as our defensive strategy, our testing procedure consisted of two steps: \begin{enumerate}
    \item Evaluate the model accuracy on the clean test data (without backdoor triggers).
    In this stage, we are merely verifying that the model performs reasonably well on clean data; we are not attempting to obtain state-of-the-art performance.
    
    \item For each adversary, construct an adversarial test set by applying the corresponding backdoor trigger to the entire clean testing dataset, and evaluate the model accuracy on the adversarial test data.
\end{enumerate}
Combining both stages, a backdoor attack is considered successful if it can force the model to misclassify most or all of the adversarial test data as the target class, while also allowing the model to correctly classify the clean test data when the backdoor trigger is absent, thus demonstrating that the misclassification is solely due to the presence of the backdoor trigger.
All experiments were repeated over 5 runs, and we report the average metrics along with the 95\% confidence intervals calculated using the two-sided Student's $t$-test.

\begin{figure}[t]
    \centering
    \includegraphics[width=\linewidth]{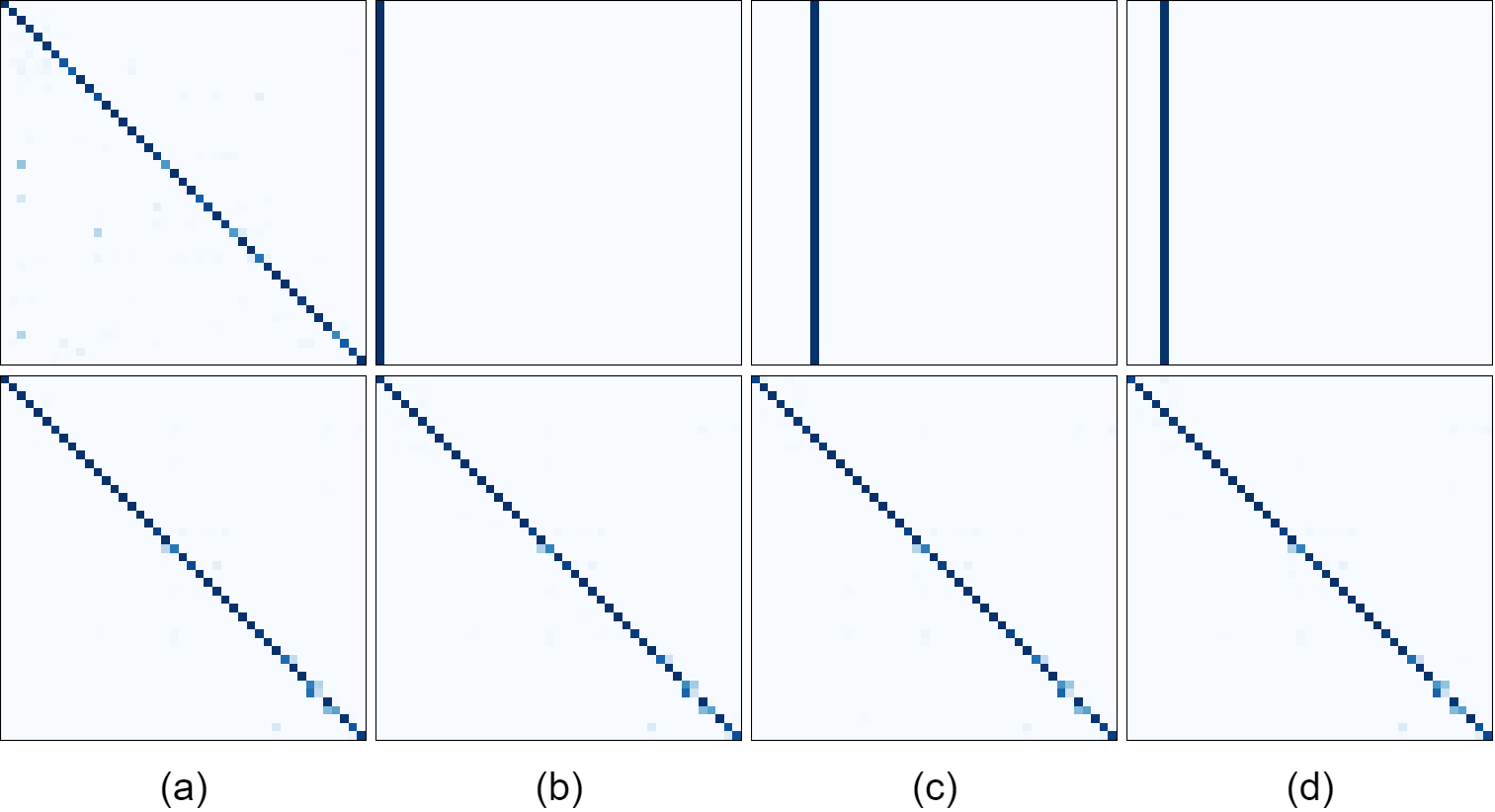}
    \caption{
        Confusion matrices for accuracy performance on the GTSRB dataset. 
        The four columns correspond to test data: (a) with clean labels, (b) with a backdoor trigger targeting class 0, (c) with a backdoor trigger targeting class 7, and (d) with a backdoor trigger targeting class 4. 
        \textbf{Top:} Confusion matrices produced by a contributor-agnostic PreAct ResNet-18. 
        While the performance of the classifier on clean data is strong, any adversary may force the classifier to produce their desired label by applying their backdoor trigger. 
        \textbf{Bottom:} Confusion matrices produced by contributor-aware training. 
        The backdoor triggers from all adversaries have been rendered ineffective. 
        }
    \label{fig:confusion}
\end{figure}

\begin{table*}[t]
    \centering
    \caption{
        Test accuracy against multiple simultaneous adversaries.
        Reported as mean and 95\% confidence interval over 5 runs.
    }
    \begin{tabular}{lccccccc}
        \multicolumn{8}{c}{\textbf{Baseline classifier}} \\ \toprule
            & \multicolumn{3}{c}{Two adversaries}   & \multicolumn{4}{c}{Three adversaries} \\ \cmidrule(lr){2-4} \cmidrule(lr){5-8}
        Dataset &   Clean   & Adv. 1    & Adv. 2    & Clean & Adv. 1    & Adv. 2    & Adv. 3 \\ \cmidrule(lr) {1-1} \cmidrule(lr){2-4} \cmidrule(lr){5-8}
        MNIST                    & $99.3 \pm 0.2$ & $9.8 \pm 0.0$  & $10.9 \pm 1.7$ & $99.1 \pm 0.2$ & $9.8 \pm 0.0$  & $10.2 \pm 0.0$ & $9.8 \pm 4.8$  \\
        CIFAR-10                 & $85.9 \pm 1.2$ & $12.1 \pm 0.4$ & $8.5 \pm 1.4$  & $84.0 \pm 2.9$ & $11.5 \pm 0.4$ & $6.5 \pm 1.6$  & $6.7 \pm 2.4$  \\
        GTSRB                    & $91.2 \pm 1.5$ & $0.4 \pm 0.0$  & $3.5 \pm 0.0$  & $93.9 \pm 1.4$ & $0.7 \pm 0.0$  & $3.5 \pm 0.0$  & $5.2 \pm 0.0$  \\ \bottomrule \\
        \multicolumn{8}{c}{\textbf{Contributor-aware training}}                                                                                              \\ \toprule
                                 & \multicolumn{3}{c}{Two adversaries}                    & \multicolumn{4}{c}{Three adversaries}                                     \\ \cmidrule(lr){2-4} \cmidrule(lr){5-8}
        Dataset                  & Clean            & Adv. 1       & Adv. 2       & Clean            & Adv. 1       & Adv. 2       & Adv. 3       \\ \cmidrule(lr) {1-1} \cmidrule(lr){2-4} \cmidrule(lr){5-8}
        MNIST                    & $97.6 \pm 0.1$ & $97.2 \pm 0.2$ & $97.3 \pm 0.1$ & $95.1 \pm 5.8$ & $93.9 \pm 6.5$ & $95.7 \pm 1.8$ & $95.0 \pm 2.4$ \\
        CIFAR-10               & $93.4 \pm 0.1$ & $93.5 \pm 0.1$ & $93.4 \pm 0.2$ & $92.6 \pm 0.2$ & $92.6 \pm 0.3$ & $92.6 \pm 0.2$ & $92.5 \pm 0.2$ \\
        GTSRB                    & $97.8 \pm 0.2$ & $97.4 \pm 0.2$ & $97.4 \pm 0.2$ & $97.5 \pm 0.2$ & $96.9 \pm 0.2$ & $97.0 \pm 0.2$ & $97.0 \pm 0.3$ \\ \bottomrule
    \end{tabular}
    \label{tab:multiple-adversaries}
\end{table*}

\subsection{Baseline classifier}
\label{sec:baseline}
For all datasets, the baseline classifier was a randomly-initialized PreAct ResNet-18, which was trained for 30 epochs.
The optimizer was stochastic gradient descent (SGD), with a learning rate of 0.02, momentum of 0.9, and weight decay of \num{5e-4}.
For the GTSRB dataset, the training and testing images were resized to a standard size of $32 \times 32$, and the learning rate was annealed by a factor of 0.1 at epochs 15 and 25.
For both the CIFAR-10 and GTSRB datasets, the training images were augmented using RandAugment \cite{cubuk2020randaugment}.

The experiments on the baseline classifier verified that the backdoor attack was effective with pinpoint precision: While the samples in the clean test dataset were classified correctly with expected performance, the introduction of the backdoor trigger into the test dataset caused nearly universal misclassification of all test samples into the classes targeted by each adversary (Figure \ref{fig:confusion}).

\subsection{MNIST experiments}
For the MNIST experiments, adversaries injected a backdoor pattern into one of the corners of the sample.
For the semi-supervised predictive ensemble, we trained one randomly-initialized auxiliary deep generative model (ADGM) \cite{maaloe2016auxiliary} for each of the five subsets of the training data (corresponding to the five contributors to the dataset). 
After the ADGMs were trained, they were independently tested on the clean testing data as well as the malicious test data with backdoor triggers. 
The outputs of each ADGM were integrated into a single label per instance using OpinionRank \cite{dawson2021opinionrank}.

\subsection{CIFAR-10 experiments}
For the CIFAR-10 experiments, adversaries injected a backdoor pattern into one of the corners of the sample.
We trained one randomly-initialized Wide ResNet-28 for each of the five subsets of the training data using the FixMatch algorithm \cite{sohn2020fixmatch}.
After all FixMatch models were trained, they were independently tested on the clean testing data as well as the malicious test data with backdoor triggers. 
The outputs of each FixMatch model were integrated into a single label per instance using OpinionRank.

\subsection{GTSRB experiments}
For the GTSRB experiments, adversaries placed a backdoor pattern at a randomly-chosen location within the region of interest of the sample (ground truth for this region is provided by the dataset). 
This strategy most closely represents our updated threat model: for example, a malicious actor wishing to attack autonomous vehicles could place stickers featuring the backdoor patterns on physical street signs.
We trained one randomly-initialized Wide ResNet-28 for each of the five subsets of the training data using the FixMatch algorithm.
The outputs of each FixMatch model were integrated into a single label per instance using OpinionRank.

\subsection{Analysis of results}
Table \ref{tab:1-adv} shows the performances of both the baseline classifier, as well as that of contributor-aware training, against a single adversary, on all three datasets.
As discussed in Section \ref{sec:baseline}, the baseline classifier exhibited a catastrophic failure in accuracy as the adversary forced the classifier to produce the target output. 
Even worse, the baseline classifier was vulnerable not only to a single adversary, but to multiple simultaneous adversaries: Table \ref{tab:multiple-adversaries} shows how \textit{any} adversary who contributes to the training database is able to compromise a na\"ive classifier. 
Note that the test examples from the target classes were not removed from the test set, so the baseline classifier retains a lower bound for performance corresponding to the intersection of the target class columns with the diagonal of the confusion matrix (see Figure \ref{fig:confusion}).

In contrast, both Table \ref{tab:1-adv} and Table \ref{tab:multiple-adversaries} show that contributor-aware training produces models that are robust to adversarial backdoor triggers.
In the single-adversary scenario, even though the adversary contributed 10\% of the training database, the model was still able to produce correct classifications in the presence of the adversary's backdoor trigger, exhibiting no meaningful change in performance.
Even against multiple simultaneous adversaries, the models produced by contributor-aware training remain resilient against all backdoor attacks. 

Figure \ref{fig:confusion} highlights the improvement of contributor-aware training over the contributor-agnostic  baseline classifier.
The baseline classifier is vulnerable to all adversaries, and is forced to produce each adversary's desired target output.
In contrast, the contributor-aware model is completely robust against all adversaries, with the model's performance unaffected by the presence of any backdoor triggers.

\section{Conclusion}
Backdoor poisoning attacks on computer vision systems reveal potential vulnerabilities in sensitive applications.
In this work, we discussed previous adversarial threat models, and proposed a new, more practical recontextualization of the problem formulation that considers the data collection process as part of the training pipeline.
In particular, we propose that a diligent data collector will trivially retain (anonymous) associations between contributions to a training database and the (potentially adversarial) users who make these contributions.
The contributor awareness granted by this recontextualization opens new avenues for defenses against backdoor attacks---crucially, \textit{without} the need to perform adversary identification or backdoor pattern detection.
Under this framework, we developed a novel universal defensive strategy that produces a classifier that is robust against backdoor attacks, while remaining agnostic to the potential adversarial characteristics of the training or testing data.
Furthermore, we showed how our defensive strategy remains robust even in the presence of multiple simultaneous adversaries, which is a novel threat that has not been previously explored.

While previous approaches toward backdoor defenses have attempted to perform careful data sanitization or model sensitivity analysis, our method requires no special consideration of any particular training or testing samples, and explicitly allows the membership of compromised models within the predictive ensemble.
Despite this relaxation of traditional constraints, we have shown that our defensive strategy produces ensembles that are resilient to the presence of backdoor patterns, even from multiple simultaneous adversaries.
Thus, our procedure represents a training pipeline that is \textit{inherently robust} to backdoor attacks, and which can be used in any computer vision application as a plug-and-play framework for secure learning.
We hope that the proposed addition of contributor awareness to the data-gathering stage of the training pipeline becomes universally standard, as such awareness opens up new dimensions of defenses against adversarial attacks.




\section*{Acknowledgment}

This work was supported by the U.S. Department of Education, GAANN Grant No. P200A180055.

\ifCLASSOPTIONcaptionsoff
  \newpage
\fi



%
\bibliographystyle{ieee_fullname}
\bibliography{1_bib.bib}

\begin{thebibliography}{10}
\providecommand{\url}[1]{#1}
\csname url@samestyle\endcsname
\providecommand{\newblock}{\relax}
\providecommand{\bibinfo}[2]{#2}
\providecommand{\BIBentrySTDinterwordspacing}{\spaceskip=0pt\relax}
\providecommand{\BIBentryALTinterwordstretchfactor}{4}
\providecommand{\BIBentryALTinterwordspacing}{\spaceskip=\fontdimen2\font plus
\BIBentryALTinterwordstretchfactor\fontdimen3\font minus
  \fontdimen4\font\relax}
\providecommand{\BIBforeignlanguage}[2]{{%
\expandafter\ifx\csname l@#1\endcsname\relax
\typeout{** WARNING: IEEEtran.bst: No hyphenation pattern has been}%
\typeout{** loaded for the language `#1'. Using the pattern for}%
\typeout{** the default language instead.}%
\else
\language=\csname l@#1\endcsname
\fi
#2}}
\providecommand{\BIBdecl}{\relax}
\BIBdecl

\bibitem{ozbayoglu2020deep}
A.~M. Ozbayoglu, M.~U. Gudelek, and O.~B. Sezer, ``Deep learning for financial
  applications: A survey,'' \emph{Appl. Soft Comput.}, vol.~93, p. 106384,
  2020.

\bibitem{maqueda2018event}
A.~I. Maqueda, A.~Loquercio, G.~Gallego, N.~Garc{\'\i}a, and D.~Scaramuzza,
  ``Event-based vision meets deep learning on steering prediction for
  self-driving cars,'' in \emph{CVPR}, 2018, pp. 5419--5427.

\bibitem{leibig2017leveraging}
C.~Leibig, V.~Allken, M.~S. Ayhan, P.~Berens, and S.~Wahl, ``Leveraging
  uncertainty information from deep neural networks for disease detection,''
  \emph{Sci. Rep.}, vol.~7, no.~1, pp. 1--14, 2017.

\bibitem{43405}
\BIBentryALTinterwordspacing
I.~Goodfellow, J.~Shlens, and C.~Szegedy, ``Explaining and harnessing
  adversarial examples,'' in \emph{ICLR}, 2015. [Online]. Available:
  \url{http://arxiv.org/abs/1412.6572}
\BIBentrySTDinterwordspacing

\bibitem{carlini2017towards}
N.~Carlini and D.~Wagner, ``Towards evaluating the robustness of neural
  networks,'' in \emph{IEEE Secur Priv}.\hskip 1em plus 0.5em minus 0.4em\relax
  IEEE, 2017, pp. 39--57.

\bibitem{NEURIPS2018_22722a34}
A.~Shafahi, W.~R. Huang, M.~Najibi, O.~Suciu, C.~Studer, T.~Dumitras, and
  T.~Goldstein, ``Poison frogs! targeted clean-label poisoning attacks on
  neural networks,'' in \emph{NeurIPS}, vol.~31, 2018.

\bibitem{gu2019badnets}
T.~Gu, K.~Liu, B.~Dolan-Gavitt, and S.~Garg, ``Badnets: Evaluating backdooring
  attacks on deep neural networks,'' \emph{IEEE Access}, vol.~7, pp.
  47\,230--47\,244, 2019.

\bibitem{tran2018spectral}
B.~Tran, J.~Li, and A.~Madry, ``Spectral signatures in backdoor attacks,'' in
  \emph{NeurIPS}, 2018, pp. 8011--8021.

\bibitem{chen2019detecting}
B.~Chen, W.~Carvalho, N.~Baracaldo, H.~Ludwig, B.~Edwards, T.~Lee, I.~Molloy,
  and B.~Srivastava, ``Detecting backdoor attacks on deep neural networks by
  activation clustering,'' in \emph{SafeAI@ AAAI}, 2019.

\bibitem{wang2019neural}
B.~Wang, Y.~Yao, S.~Shan, H.~Li, B.~Viswanath, H.~Zheng, and B.~Y. Zhao,
  ``Neural cleanse: Identifying and mitigating backdoor attacks in neural
  networks,'' in \emph{IEEE Secur Priv}.\hskip 1em plus 0.5em minus 0.4em\relax
  IEEE Computer Society, 2019, pp. 707--723.

\bibitem{Wenger_2021_CVPR}
E.~Wenger, J.~Passananti, A.~N. Bhagoji, Y.~Yao, H.~Zheng, and B.~Y. Zhao,
  ``Backdoor attacks against deep learning systems in the physical world,'' in
  \emph{CVPR}, June 2021, pp. 6206--6215.

\bibitem{NEURIPS2020_0ea6f098}
A.~Pal and R.~Vidal, ``A game theoretic analysis of additive adversarial
  attacks and defenses,'' in \emph{NeurIPS}, vol.~33, 2020, pp. 1345--1355.

\bibitem{nguyen2020input}
T.~A. Nguyen and A.~Tran, ``Input-aware dynamic backdoor attack,''
  \emph{NeurIPS}, vol.~33, pp. 3454--3464, 2020.

\bibitem{chan2019poison}
A.~Chan and Y.-S. Ong, ``Poison as a cure: Detecting \& neutralizing
  variable-sized backdoor attacks in deep neural networks,'' \emph{arXiv
  preprint arXiv:1911.08040}, 2019.

\bibitem{gao2019strip}
Y.~Gao, C.~Xu, D.~Wang, S.~Chen, D.~C. Ranasinghe, and S.~Nepal, ``Strip: A
  defence against trojan attacks on deep neural networks,'' in \emph{Annu.
  Comput. Secur. Appl. Conf.}, 2019, pp. 113--125.

\bibitem{udeshi2019model}
S.~Udeshi, S.~Peng, G.~Woo, L.~Loh, L.~Rawshan, and S.~Chattopadhyay, ``Model
  agnostic defence against backdoor attacks in machine learning,'' \emph{arXiv
  preprint arXiv:1908.02203}, 2019.

\bibitem{doan2020februus}
B.~G. Doan, E.~Abbasnejad, and D.~C. Ranasinghe, ``Februus: Input purification
  defense against trojan attacks on deep neural network systems,'' in
  \emph{Annu. Comput. Secur. Appl. Conf.}, 2020, pp. 897--912.

\bibitem{liu2018fine}
K.~Liu, B.~Dolan-Gavitt, and S.~Garg, ``Fine-pruning: Defending against
  backdooring attacks on deep neural networks,'' in \emph{RAID}.\hskip 1em plus
  0.5em minus 0.4em\relax Springer Verlag, 2018, pp. 273--294.

\bibitem{liu2019abs}
Y.~Liu and W.-C. Lee, ``Abs: Scanning neural networks for back-doors by
  artificial brain stimulation,'' in \emph{CCS}, 2019.

\bibitem{kolouri2020universal}
S.~Kolouri, A.~Saha, H.~Pirsiavash, and H.~Hoffmann, ``Universal litmus
  patterns: Revealing backdoor attacks in cnns,'' in \emph{CVPR}.\hskip 1em
  plus 0.5em minus 0.4em\relax IEEE, 2020, pp. 298--307.

\bibitem{nguyen2021wanet}
\BIBentryALTinterwordspacing
T.~A. Nguyen and A.~T. Tran, ``Wanet - imperceptible warping-based backdoor
  attack,'' in \emph{ICLR}, 2021. [Online]. Available:
  \url{https://openreview.net/forum?id=eEn8KTtJOx}
\BIBentrySTDinterwordspacing

\bibitem{Doan_2021_ICCV}
K.~Doan, Y.~Lao, W.~Zhao, and P.~Li, ``Lira: Learnable, imperceptible and
  robust backdoor attacks,'' in \emph{ICCV}, October 2021, pp.
  11\,966--11\,976.

\bibitem{chen2017targeted}
X.~Chen, C.~Liu, B.~Li, K.~Lu, and D.~Song, ``Targeted backdoor attacks on deep
  learning systems using data poisoning,'' \emph{arXiv preprint
  arXiv:1712.05526}, 2017.

\bibitem{barni2019new}
M.~Barni, K.~Kallas, and B.~Tondi, ``A new backdoor attack in cnns by training
  set corruption without label poisoning,'' in \emph{ICIP}.\hskip 1em plus
  0.5em minus 0.4em\relax IEEE, 2019, pp. 101--105.

\bibitem{tao2020label}
F.~Tao, L.~Jiang, and C.~Li, ``Label similarity-based weighted soft majority
  voting and pairing for crowdsourcing,'' \emph{KAIS}, vol.~62, pp. 2521--2538,
  2020.

\bibitem{dawson2021opinionrank}
G.~Dawson and R.~Polikar, ``Opinionrank: Extracting ground truth labels from
  unreliable expert opinions with graph-based spectral ranking,'' in
  \emph{IJCNN}, 2021, pp. 1--8.

\bibitem{lecun2010mnist}
Y.~LeCun, C.~Cortes, and C.~Burges, ``Mnist handwritten digit database,''
  \emph{ATT Labs [Online]. Available: http://yann.lecun.com/exdb/mnist},
  vol.~2, 2010.

\bibitem{krizhevsky2009learning}
A.~Krizhevsky, G.~Hinton \emph{et~al.}, ``Learning multiple layers of features
  from tiny images,'' 2009.

\bibitem{Stallkamp2012}
J.~Stallkamp, M.~Schlipsing, J.~Salmen, and C.~Igel, ``Man vs. computer:
  Benchmarking machine learning algorithms for traffic sign recognition,''
  \emph{Neural Netw}, vol.~32, pp. 323--332, 2012.

\bibitem{raji2020saving}
I.~D. Raji, T.~Gebru, M.~Mitchell, J.~Buolamwini, J.~Lee, and E.~Denton,
  ``Saving face: Investigating the ethical concerns of facial recognition
  auditing,'' in \emph{AAAI/ACM Conf. AI Ethics Soc.}, ser. AIES '20, 2020, p.
  145–151.

\bibitem{cubuk2020randaugment}
E.~D. Cubuk, B.~Zoph, J.~Shlens, and Q.~V. Le, ``Randaugment: Practical
  automated data augmentation with a reduced search space,'' in \emph{CVPR},
  2020, pp. 702--703.

\bibitem{maaloe2016auxiliary}
L.~Maal{\o}e, C.~K. S{\o}nderby, S.~K. S{\o}nderby, and O.~Winther, ``Auxiliary
  deep generative models,'' in \emph{ICML}.\hskip 1em plus 0.5em minus
  0.4em\relax PMLR, 2016, pp. 1445--1453.

\bibitem{sohn2020fixmatch}
K.~Sohn, D.~Berthelot, N.~Carlini, Z.~Zhang, H.~Zhang, C.~A. Raffel, E.~D.
  Cubuk, A.~Kurakin, and C.-L. Li, ``Fixmatch: Simplifying semi-supervised
  learning with consistency and confidence,'' \emph{NeurIPS}, vol.~33, 2020.

\end{thebibliography}

%








\end{document}